\documentclass[11pt]{article}
\usepackage{moriond,epsfig,amsmath,amssymb}

\begin{document}
\vspace*{4cm}
\title{EVOLUTION OF RANDALL-SUNDRUM BLACK HOLES IN DENSE MEDIA}
\author{V. VAN ELEWYCK}
\address{Service de Physique Th\'eorique, Universit\'e Libre de Bruxelles - CP 225, Boulevard du Triomphe, 1050 Bruxelles, Belgium}
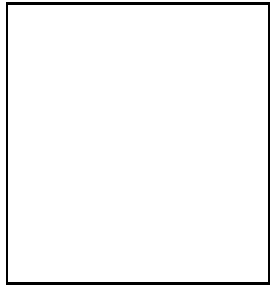
\maketitle\abstracts{In the framework of Randall-Sundrum model, we discuss the evolution of microscopic black
  holes created by cosmic particles hitting a neutron or a
  strange star. We set up the general equations of evolution for the black hole mass and momentum in a dense medium, and discuss the possibility
  of collapse due to the black hole's growth inside the star.}

\section{Introduction}
In models with extra space dimensions, the energy threshold for microscopic black hole (MBH) production is only slightly higher than the fundamental energy scale in (4+n) dimensions, $M_F$, which can be as low as a TeV. MBH produced at accelerators or in cosmic ray showers will immediately decay due to Hawking radiation; but they may survive longer if produced in a very dense medium, where evaporation and accretion of matter from the surroundings are two competing processes. Such a situation can arise when an ultra-high energy cosmic particle hits the matter inside a neutron or a strange star. In our paper\cite{US}, we work out the general equations of evolution for a MBH in those two cases, and investigate whether the black hole can grow fast enough to engulf the whole star, as was suggested in the context of flat extra dimensions\cite{LEA}. 
We have addressed the problem in the framework of Randall-Sundrum model\cite{RS} (with one, warped, extra dimension), for which few phenomenological constraints have been derived until now. The evolution of the MBH here depends on the parameter $\mu$, the inverse curvature radius of the slice of $AdS_5$ as seen from our brane.

\section{Matter with $T\ll m_{target}$ : neutron stars}

A MBH created inside a neutron star moves through a medium of nucleons of mass $m$ and number density $n$, at $T=0$. The evolution of the MBH mass $M_B$ depends on the accretion rate of target particles, and on the mass losses due to Hawking evaporation:\begin{equation}
 \frac{dM_B}{dt} =\left.\frac{dM_B}{dt}\right|_{acc}+\left.\frac{dM_B}{dt}\right|_{evap} = \pi r_s^2\beta m\left\{n-\frac{g_{eff}}{960\pi^2 m}\frac{M_B}{pr_s^4}\right\}\label{massneutron}
\end{equation}

where $\beta = P_B/E$, $\gamma = E/M_B$ and $r_s$ is the Schwarzschild radius of the MBH. As a consequence of the isotropy of Hawking radiation in the black hole rest frame, we also have to consider a variation of momentum:
\begin{equation}
\frac{dP_B}{dt}=\frac{p}{M_B}\frac{1}{\gamma}\left.\frac{dM_B}{d\tau}\right|_{evap}
\label{momneutron}
\end{equation}

The critical density of target particles for keeping the MBH growing (\noindent$n_{crit} \simeq 10^{15} gr/cm^3$), is however much higher than the neutron star's crustal density, where the cosmic neutrino will interact. We thus expect the black hole to rapidly decay.

\section{Matter with $T\gg m_{target}$ : strange stars}

In a bare strange star, the condensate of u, d and s quarks forms a medium of relativistic particles at a temperature $T_{bath}\simeq 1$ GeV, which in the MBH rest frame corresponds to an effective temperature\cite{LEA} $T_{eff}= T_{bath}\left( 1+\frac{4}{3}\frac{P_B^2}{M_B^2}\right)^{1/4}$. The mass evolution (in the MBH rest frame) is now given by:
\begin{equation}
\frac{dM_{B}}{d\tau}=\frac{\pi^3 g_{eff}}{15}r_s^2\left(T_{eff}^4-\frac{1}{4(2\pi)^4 r_s^4}\right).
\label{tgrow}
\end{equation}
Here $T_{crit}\simeq 100$ GeV $\ll T_{eff}$ so that the MBH can keep on growing, travelling across the star.

\section{Conclusions on the possibility of star collapse due to black hole seeding}
The MBH created in a strange star rapidly reaches the size $\mu^{-1}$ and begins to feel the warping of space. Its radius then increases much slowlier with mass (we assume\cite{BHlog} $r_s\sim\frac{1}{\mu}\ln\left(\frac{\mu^2M_B}{M_F^3}\right)$), so that the Hawking evaporation rate stays high and the MBH looses momentum, seeing a medium with lower and lower $T_{eff}$. Numerical analysis showed that the black hole either slows down and decay inside the star, or crosses it and escapes before it could accrete enough matter to engulf the star.
An alternative scenario\cite{SUR}, in which entropy causes the MBH to split up into smaller black holes, leads to the same conclusion: they will all slow down and decay unless $T_{bath}$ is higher than the minimum black hole temperature ($\sim \mu$). This condition translates into an upper bound of $\mu\leq 20$ GeV, a region of parameter space already ruled out by collider constraints\cite{DIA}. We thus exclude the possibility of neutron and strange star collapse driven by a RS black hole.

\section{Acknowledgements}
V. Van Elewyck acknowledges financial support from I.I.S.N. and I.A.P. Program. She thanks her collaborator Malcolm Fairbairn, and is grateful to the organizers for their hospitality.

\end{document}